\documentclass[11pt,a4paper,notitlepage]{article}
\usepackage{amsfonts,amssymb} 
\usepackage{graphics} 

\begin{document}
\pagenumbering{arabic}\setcounter{page}{1}
\thispagestyle{empty}
\begin{titlepage}

\begin{flushleft}

{\bf {\large  Quantum Information Processing in Nanostructures}}\\

\vskip3cm

{\bf Alexandra Olaya-Castro}\\
Centre for Quantum Computation, Clarendon Laboratory\\
Department of Physics, University of Oxford\\
Parks Road, Oxford OX1 3PU, United Kingdom\\
\vskip1cm
E-mail: a.olaya@physics.ox.ac.uk\\
Phone: 44 1865 272308\\
Fax: 44 1865 272400\\
\vskip2cm  
{\bf Neil F. Johnson}\\
Centre for Quantum Computation, Clarendon Laboratory\\ 
Department of Physics, University of Oxford\\
Parks Road, Oxford OX1 3PU, United Kingdom\\
\vskip1cm
E-mail:n.johnson@physics.ox.ac.uk\\
Phone:44 1865 272287\\
Fax: 44 1865 272400

\vskip0.5cm
to appear in {\em ``Handbook of Theoretical and Computational Nanotechnology"}
\end{flushleft}

%\maketitle

\end{titlepage}

\tableofcontents
\pagenumbering{arabic}\setcounter{page}{1}
\setcounter{section}{0}\newpage
\section{Introduction}
%\markboth{Introduction}{}
%\noindent
%\subsection{What for quantum information processing systems?}

What can be accomplished in a world governed by quantum correlations? 
This question has brought together researchers from a wide range of backgrounds -- from fundamental physics through to information technology -- in the common pursuit of understanding, designing and eventually building quantum information processing (QIP) systems. 

The fundamental unit of quantum information is the quantum bit, or {\em qubit}, which is a quantum version of the classical binary bit (i.e. 0 and 1). Thanks to the enormous amount of work in the QIP field over the past decade, we now know that it should be possible to use qubits to realize fundamentally new and more powerful methods for computation and communication. In other words, one ought to be able to harness the `spookiness' of quantum mechanics, in particular quantum correlations, to achieve a revolutionary form of information processing. It has been shown, for example, that algorithms for factorizing large 
integers can be realized faster in a suitably chosen QIP system than in a conventional computer\cite{shor97}. The teleportation of a quantum state  between spatially distinct locations, has been experimentally demonstrated\cite{bouwmeester97, boschi98}. An application of increasing importance in this field, concerns the use of QIP to simulate other quantum systems\cite{somaroo99, tseng00,vidalep03}. Interesting by-products of this research include a new understanding of physical behaviour at a quantum phase transition\cite{osterloh02,osborne02, vidal03}. The promise of further exciting applications and a deeper fundamental understanding of quantum mechanics, has sparked off a very active QIP research community spanning physicists, mathematicians, materials scientists, chemists and computer scientists.

\subsection{The Challenge of Scalability}
Any practical implementation of a QIP system needs to meet several stringent criteria in order to operate successfully\cite{divincenzofp00}. First of all, the qubits (which can also be thought of as `quantum memories') must be sufficiently isolated so that they can then be directly and conditionally manipulated in
 a controlled environment. They need to be initialized precisely and then 
efficiently measured. The effective
interactions among qubits should also be carefully tuned, and a set of 
universal quantum operations should be made possible in order to perform any other required quantum gate. Most importantly, the system must be scalable to
more than a few qubits.  
Consequently a large-scale QIP system is expected to include  
of the order of tens to hundreds of qubits arranged in a configurable way, depending on the quantum routine to be achieved. In order to maintain the quantum correlations, parallel addressing of spatially separated units is also required, with operation times smaller than both local
and non-local decoherence times. Scalability and robustness are hence arguably two of the most demanding requirements facing the practical implementation of quantum information processing.

\subsection{Nanostructures for scalable QIP}
The clearest demonstrations of QIP involving massive particles,
have come from the field of interacting atoms/ions and
photons\cite{monroe02}. Individual atoms or ions, which can be
strongly or weakly coupled to the photon field, are manipulated in controlled
environments with well-understood couplings and well-defined decay
channels, thereby offering an ideal experimental set-up to
systematically study the basic principles of quantum computation and
communication. In fact, trapped ions in optical lattices have become
an ideal test ground for the study of more complex QIP
architectures\cite{blatt03,blatt203}. Despite these achievements, there is a growing feeling within the QIP community that the
potential integration of basic science and technology which will be required  for building large-scale quantum processors, will be best achieved in solid-state systems. After all, the transistor gave us classical computing, so why shouldn't a quantum version give us quantum computing?
 
In the last decade, there have been a number of proposals of solid state systems in which quantum
information processing might be achieved experimentally\cite{loss98,kane98,quiroga99, biolatti00, platzman99, makhlin01}. All these proposals are conceptually similar in the
sense that they require 
manipulation of the coherent dynamics of a quantum system which is 
embedded in a complex environment. Hence they all share the same challenge 
of having to resist highly non-trivial decoherence sources. Given that decoherence times in solid-state systems are short, the structures themselves must be sufficiently small that the quantum information can be manipulated and/or transferred in a time which is shorter than the decoherence time. Remarkably, such small structures, or {\em nanostructures}, are now beginning to be built. In fact, the incredible advances in solid-state nanofabrication mean that artificial and natural nanostructures, such as quantum dots, are becoming prime candidates for building a robust large-scale QIP system. 

Semiconductor quantum dots (QDs) \cite{bimberg, harrison} are structures  with the dimensions of the order
of few nanometers. Their electronic and optical properties can be tailored
through the quantum confinement which results from their finite size. In particular, they exhibit physical features similar to
individual atoms but are typically a few orders of magnitude larger and have an energy scale which is a few orders of magnitude
smaller. Because of their atom-like properties\cite{bimberg} but their key
difference in size, they can be regarded as localized and addressable
units for storing and manipulating  quantum information. Quantum information can be
encoded in QDs through a variety of effective two-level systems -- for example, electron
spin or charge excitation. Even more interesting is the possibility of building an artificial
`molecule' by coupling together two QDs. As we will discuss later,
the nature of the interaction between QDs not only offers different
perspectives for QIP, but also establishes a direct link with
biological and organic systems.

Although the earliest proposals for QIP with QDs envisioned spin
interactions controlled by electrical gates\cite{loss98,bukard99}, it is now thought 
that schemes based on all-optical control of the quantum
system and its interactions are possibly more desirable since they offer
two principal advantages. First, the use of ultrafast laser
technology\cite{shah} means that quantum operations can be carried out
within the  
coherence time. Second, the interaction between a QD (acting as a stationary quantum memory) and photons (acting as a channel for information transmission) make this setup attractive for both the manipulation and
transfer of quantum information over relatively large distances. In this latter context,
QDs can also be integrated with optical nano-cavities offering a means for
exploiting the techniques of cavity quantum electrodynamics\cite{imamoglu99, kiraz03}.

With this backdrop, the aim of this Chapter is to survey the current state of theoretical proposals of QIP in semiconductor QDs, focusing on schemes which exploit optical control of qubit interactions. Readers interested in other solid state
proposal are invited to consult Ref.\cite{reviewsolid}. The Chapter is organized as follows. General theoretical aspects of the field of QIP are reviewed in Section II. Theoretical proposals for all-optical QIP using QDs are reviewed in Section III. Finally, future trends and developments are discussed in Section IV.

\setcounter{section}{1}\newpage
\section{Theoretical Background}
The coherent superposition and entanglement of quantum states are the features
lying at the heart of any QIP protocol. They account for quantum
interference effects within a single system (superposition) or
between different quantum systems (entanglement). In any
experimental situation, however, the unavoidable interaction of a
quantum system with its environment leads to decoherence processes
which eventually destroy superpositions and entanglement and yield
statistical mixtures of states. Understanding such complex decoherence
mechanisms and finding approaches to make quantum
interference effects robust in the face of such decoherence, are still the major challenges facing many QIP proposals.

\subsection{Basic QIP toolbox: Superposition and Entanglement}
{\em Quantum coherence} represents the ability of a quantum system
to be in a superposition of different states, and is the crucial element that
leads to the definition of the fundamental unit of quantum
information: the quantum bit (qubit).  Qubits are quantum
systems that can be represented to a good approximation as a
two-state system. As a consequence, the state of an isolated qubit
can be expressed as a {\em coherent superposition} of these two
states:
\begin{equation}
|\Phi\rangle = \alpha|0\rangle + \beta |1\rangle \nonumber
\label{Eq:qubit}
\end{equation}
where $|\alpha|^2$ and $|\beta|^2$ are the probabilities that the system
be found in state $|0\rangle$ or $|1\rangle$, respectively.
Such a coherent superposition implies that there is always a basis
in which the value of the qubit is well-defined, as opposed to an
incoherent mixture in which the qubit's state becomes a
statistical mixture irrespective of the basis used\cite{bouwmeester}. 

A single two-state quantum system can be
represented as an effective spin$-\frac{1}{2}$ particle in a local
magnetic field\cite{rose}, whose Hamiltonian can be written as
\begin{eqnarray}
\mathcal {H}_0(t)=B(t)\cdot \sigma
\end{eqnarray}
Here $\sigma_{x,y,z}$ are the Pauli matrices describing the qubit
to be manipulated. Full control over the coherent dynamics of the
system is possible if at least two components of the effective
field $B(t)$ can be switched arbitrarily. This Hamiltonian
corresponds to an ideal physical scheme for realizing single-qubit
unitary operations described by the operator $U(t,0)=\mathcal {T}
exp(\int_{0}^t \mathcal {H}_0(t')dt')$ ($\mathcal {T}$ denotes
time-ordering). For instance, under the action of $U(t,0)$ it is
possible to perform a Hadamard gate\cite{nielsen} which transforms
the states $|0\rangle$ and $|1\rangle$ into coherent
superpositions of the type described in Eq.(\ref{Eq:qubit}) as
$H|0\rangle \mapsto (1/\sqrt 2)(|0\rangle + |1\rangle)$ and $H|1\rangle \mapsto (1/\sqrt 2)(|0\rangle - |1\rangle)$.

However, any physical realization of a qubit -- for example, an electron's spin up or down, an optically excited QD with one exciton or no exciton, or the flux in a SQUID -- is embedded in a weakly interacting
environment with a large number of degrees of freedom. This yields
transient behavior with only partial coherence. Thus a density
matrix formalism is more convenient to describe the qubit dynamics,
where the reduced density operator in the basis $\{|0\rangle,
|1\rangle\}$ can be written as
\begin{eqnarray}
\hat \rho_q=\left (
\begin{array}{*{2}{c}}
\rho_{00} & \rho_{01} \\
\rho_{10} & \rho_{11}
\end{array}\right )
\end{eqnarray}
The diagonal elements $\rho_{00}$ and $\rho_{11}$ denote the
populations of each level and the off-diagonal elements
$\rho_{10}$ and $\rho_{01}$ define the coherences between states
$|0\rangle$ and $|1\rangle$. The population of the
excited level $\rho_{11}$ is typically described as having a characteristic decay time known as
$T_1$, while the typical decay time of the coherences $\rho_{10}$ and $\rho_{01}$ is known as the dephasing or coherence time $T_2$.
If, however, the qubit dynamics are explored over timescales
significantly smaller than $T_2$ using ultrafast spectroscopy, the dephasing processes may not have sufficient time to be completed and hence the effects of memory in these processes must be accounted for. The Markov approximation in which memory effects are neglected yielding a characteristic exponential decay,  may no longer be sufficiently accurate. Consequently $T_2$ may depend on the temporal correlations between the qubit system and the environment and within the environment itself. In semiconductor nanostructures such non-Markovian effects are under active experimental and theoretical investigation\cite{borri01,rodriguez02,rodriguez03}.

The processing of quantum information requires qubits to
be coherently manipulated on timescales shorter than $T_2$. From
the experimental point of view, a fundamental probe of coherent
manipulation of a qubit is the identification of Rabi
oscillations, which are produced in resonantly driven two-state
systems\cite{allen} and have no
classical analog. They correspond to a sinusoidal time-evolution
of the population difference in a two-level system for timescales 
short as compared to the coherence time.

{\em Entanglement} occurs as a result of quantum interference
among states of composite systems, giving rise to non-classical
correlations between spatially separated quantum systems.  An
essential manifestation of these non-classical correlations is
that a measurement performed in one of the subsystems determines
the state of the other even if they are significatively far apart.
For a long time this was considered a bizarre if not `spooky' feature of quantum mechanics. However the modern view of entanglement is
to see it as a fundamental nonlocal resource which can be used in any QIP
protocol. The power of this resource has been
demonstrated in quantum communication schemes performed with
entangled photons, including state teleportation\cite{bouwmeester97,
boschi98},
cryptographic key distribution\cite{bennett92}  and quantum dense
coding\cite{bennettdc92}. Entanglement of massive particles has been
experimentally achieved in cavity QED\cite{revharoche01}, ion
traps\cite{monroe00}, and superconducting qubits\cite{belkey03}.

The most general entangled state of $N$ qubits can be expressed as
a superposition of $2^N$ states with complex and time-dependent
amplitudes $a_i$ with $i=0,1, \dots,2^N-1$:
\begin{eqnarray}
|\Psi\rangle & = &a_0(t)|0_10_20_3...0_N\rangle + a_1(t)|1_10_20_3...0_N \rangle +\nonumber \\
& & \cdots + a_{2^N-1}(t)|1_11_21_3...1_N\rangle
\label{eq:entangled}
\end{eqnarray}
As long as this state cannot be factorized into a product of single qubit states, then there will be non-classical correlations which can be exploited to store more information than a comparable set of classical counterparts, and which can be exploited to perform parallel operations on different qubits. Therein lies the extraordinary potential for quantum computation.

The quantification of entanglement has become a central
problem in quantum information theory. In the case of bipartite
quantum systems, several measures have been proposed (see Ref. \cite{brub02}) including the entanglement of
formation\cite{wootters01} which has the clear
 meaning of being the asymptotic number of Bell pairs
required to prepare the state using only local unitary
transformations and classical communication. Intimately related to
this measure is the notion of concurrence\cite{wootters98}. While
the bipartite case is well understood, a general
formulation of multipartite entanglement remains an outstanding
open problem.

\subsection{Universal resources for QIP}
One of the key issues that has been addressed in QIP  science
is to determine what subsets of physical resources are capable of achieving universal
quantum computation and, more generally, how
to quantify such resources for different QIP tasks\cite{nielsen03}. The
idea of universality is to identify a set of gates that are sufficient to perform
any other quantum operation on arbitrary N qubits -- for example, in order to transform any initial separable state of N qubits into
an arbitrary entangled state of the type described in Eq.(\ref{eq:entangled}).

It is well known that certain two-qubit gates are
universal for quantum computation when assisted with arbitrary one-qubit gates\cite{barenco95}. An example is the controlled-not ({\footnotesize CNOT}) operation which transforms as
\begin{eqnarray}
|m,n\rangle \rightarrow |m,m\oplus n\rangle \quad m \; \epsilon \{0,1\}
\end{eqnarray}
where $\oplus$ denotes addition modulo 2. 
Recently it has become clearer what general class of two qubit gates
are universal. It turns out that any interaction creating entanglement
between any pair of qubits is universal for quantum computation\cite{dur01, bennett02,bremner02}. In particular, it has been demonstrated that a fixed {\em entangling} two-qubit interaction can be used to perform universal quantum
computation, when assisted by single-qubit transformations between
applications of the entangling gate $V$\cite{bremner02}. This can be expressed 
as follows
\begin{eqnarray}
U=(A_1\otimes B_1)e^{i V}(A_2 \otimes B_2)
\end{eqnarray}
where  $A_j,B_j$ are one qubit gates and $V=\sum_{X,Y,Z}\theta_{\alpha} \sigma_{\alpha} \otimes \sigma_{\alpha}$ with $\frac{-\pi}{4}~\leq ~\theta_{\alpha}~\leq\frac{-\pi}{4}$. The key point in the demonstration is that by using single-qubit gates, any gate $U$ can be transformed into $W=e^{i\phi\sigma_Z\otimes\sigma_Z}$ with $0<|\phi|<\pi/2$, which is a natural gate for implementing the {\footnotesize CNOT} operation.

\subsubsection{\em Fundamental two-qubit interactions}
The physical implementation of a two-qubit gate depends on both
the system's  network structure and the nature of the {\em
effective} interaction coupling together two qubits. The network structure
refers to whether each qubit can be coupled to any other qubit, or
whether each qubit can only be coupled to a few other qubits (e.g. nearest neighbors). For a given quantum task,
this underlying structure is related to the number of operations
required to achieve that task. This however also depends on the
microscopic nature of the interaction between qubits. For
instance, it has been discussed that a {\footnotesize CNOT} operation can be more
efficiently realized if the available interaction  is given by an
Ising Hamiltonian rather than when it is described by an isotropic
Heisenberg interaction\cite{schuch03}. The nature of the interaction
also determines whether it is possible to use recoupling to encode qubits and hence reduce the number of 
single-qubit operations\cite{lidar02}.

In a large number of proposals for QIP in solid state systems, the fundamental two-qubit gate is generated by an exchange interaction  between qubits $i$ and $j$ which has the general form
\begin{eqnarray}
\mathcal {H}_{int}(t)=\sum_{\alpha=X,Y,Z} J^{\alpha}(t)
\sigma_i^{\alpha} \cdot \sigma_j^{\alpha}
\end{eqnarray}
This interaction may arise naturally in the system, or it might be 
effectively created by manipulating local variables like individual
energy transitions or individual couplings to a common collective mode.

The isotropic Heisenberg model, which corresponds to
$J^{\alpha}(t)\equiv J(t)$,  describes the fundamental two-qubit
interaction in proposals for the implementation of spin-based
quantum computation\cite{loss98,kane98,bukard99,vrijen00}. It has
been shown that the Heisenberg interaction is sufficient to
perform universal quantum gates when a logical qubit is encoded in
two or more qubits\cite{divincen00}. These schemes
require tunability of the interaction which may prove very
difficult to achieve experimentally. As an alternative, schems for
quantum computation with an always-on Heisenberg interaction have
been proposed\cite{benjamin03}. By tuning
individual energy transitions via local or global addressing, the
effective interaction between qubits in a linear array becomes an
effective Ising interaction: $\mathcal {H}_{int}\simeq
J\sum\sigma^Z_i\sigma_{i+1}^Z$. The interesting point here is that
up to single-qubit rotations, the {\footnotesize CNOT} operation is a natural
two-qubit gate for the Ising Hamiltonian as discussed in
Ref.\cite{bremner02,schuch03} and hence it is a resource for universal quantum
computation.

The anisotropic $XXZ$ model, which corresponds to the case where
$J^X(t)=\pm J^Y(t) \neq J^Z(t)$, is the natural interaction of
electrons on a helium surface\cite{platzman99}. The scaling properties
of entanglement close to the phase transition described by this
model, have been discussed\cite{vidal03} as well as schemes for
encoded QIP\cite{lidar02}.

The $XY$ model, corresponding to the case $J^X(t)=J^Y(t), J^Z=0$,
has been intensively studied in the context of
QIP\cite{osterloh02,vidal03,schuch03,wang01}. These studies include the
generation of so-called $W$ states\cite{dur00} in a one
dimensional $XY$ model\cite{wang01} as well as scaling properties of
entanglement in the vicinity of a quantum phase transition
described by the $XY$ interaction in a transverse magnetic
field\cite{osterloh02, vidal03}. Moreover it has been shown that
the effective dipole-dipole interaction dominates the
generation of entangled states in a wide range of systems, from excitons in quantum dots\cite{quiroga99} through to both
atomic systems\cite{zheng00} and quantum-dot spins  in
cavities\cite{imamoglu99}. In addition, it has been shown that a
natural  two-qubit gate for this interaction is a simultaneous
{\footnotesize CNOT} and {\footnotesize SWAP} operation\cite{schuch03}. 
The $XY$ interaction therefore has an important potential role to play in QIP implementations.

\subsubsection{\em Cavity QED as a resource for QIP}
The nature of the photon-matter interaction in cavity quantum
electrodynamics (QED) has very important implications for the processing of
QIP\cite{mabuchi02} for several reasons. In the strong coupling regime, it is
possible to use the cavity field as an intermediary or `bus' to create
entanglement between qubits.  Several two-qubit entangling
protocols using this scheme have been proposed\cite{zheng00, yi03,olayaprl04}
with some of them being experimentally realized\cite{revharoche01}. Cavity
QED thus provides a scenario for exploring  more complex QIP
protocols in `clean' environments. In addition, qubit-cavity systems are arguably the most
promising candidates for the distribution of quantum information in a
quantum network\cite{cirac97}. For example, qubits driven by lasers and
strongly coupled to spatially separated cavities can be
entangled via photons travelling from one cavity to another\cite{cirac97}. Finally, cavity QED has well-defined decoherence
channels. Subject to high-efficiency measurements, this provides an
experimental set-up for studying the conditional evolution of open
quantum systems\cite{mabuchi02}.

In cavity-assisted entanglement, non-interacting qubits can be
prepared in a entangled state through either dispersive\cite{zheng00} or
resonant\cite{olayaprl04} interactions with a common quantized field.
Entanglement is achieved in the transitory regime where the strong
interaction between qubits and a vacuum cavity field is switched
on and off on a time scale shorter than the cavity-field decay
time and the dipole decay time. Such an interaction can be
effectively created using a chain of single qubits, for example via a stream of flying qubits which enter and leave the cavity, as well
as via a set of qubits which simultaneously interact with the field.

The most general situation is described in the interaction picture
by the Hamiltonian ($\hbar=1$)\cite{olayajob94}
\begin{eqnarray}
\mathcal {H}_c=\sum_{i=1,2}f_i(t_i,t,\tau_i)\{e^{-i\delta t}a^\dag
\sigma_i^{-}+e^{i \delta t}\sigma_i^{+}a\}
\end{eqnarray} where $\sigma_i^{+}=|1_i\rangle \langle 0_i|$,
$\sigma_i^{-}=|0_i\rangle \langle 1_i|$ with $|1_i\rangle$ and
$|0_i\rangle$ $(i=1,2)$ being the excited and ground states of the
$i$'th qubit. Here $a^\dag$ and $a$ are respectively the creation
and annihilation operators for the cavity photons and $\delta$ is
the detuning between the qubit transition frequency and the cavity
field frequency. The time-dependent coupling of the cavity field
with the $i$'th qubit, which is injected at $t_i$ and interacts
during a time $\tau_i$, is given by a time-window function,
\begin{eqnarray}
f_i(t_i,t,\tau_i)=[\Theta (t -\tau_i)-\Theta (t
-\tau_i-t_i)]\gamma_i(t)
\end{eqnarray} where $\gamma_i(t)$ is the time-dependent qubit-field
coupling strength.
For the situation in which the second qubit interacts with the cavity
just after the first did, i.e. $t_2=t_1+\tau_1$, the Hamiltonian
corresponds to  the Jaynes-Cummings interaction
for each qubit separately.  Under resonant
conditions $\delta=0$,  the initial
maximum entanglement  between a first qubit and the field is converted
into qubit-qubit entanglement during the interaction of the second
qubit with the cavity mode\cite{olayaprl04}. Einstein-Podolsky-Rosen (EPR)
atomic pairs and three particle entangled states have experimentally been achieved using this sequential scheme\cite{revharoche01}.

For the case of two qubits which simultaneously interact with the
same cavity field, we have $t_1=t_2$ and $\tau_1=\tau_2$. A far richer  control space can be explored if we generalize the setup to 
different and time-dependent qubit-cavity couplings. This latter situation
is  described by a generalized two-qubit {\em dynamical Dicke
(DD)} model. Under non-resonant conditions where $\delta>>\gamma_1\gamma_2$, the common quantized field is only virtually
excited. This gives rise to an effective interaction corresponding to
the Heisenberg $XY$ model\cite{zheng00} in which a coherent energy
transfer takes place between pairs of qubits. This scheme has been 
experimentally realized with two Rydberg atoms crossing a nonresonant cavity\cite{osnaghi01}.

When both qubits are on resonance with the cavity mode and are
simultaneously interacting, the generation of entanglement can
be controlled using asymmetric qubit-cavity
couplings\cite{olayaprl04}. This scheme exploits a trapping vacuum
condition in which the cavity-qubit state becomes separable,
leaving the cavity photon number unchanged but the two-qubit subsystem
in an entangled state. Such a trapping condition does not arise for
identical couplings.

\subsection{General aspects of decoherence}
Decoherence is generally a fast process whose timescale depends
primarily on the size and temperature of the system, but may also depend
on other factors as well (e.g. sample preparation, errors in the prescribed unitary evolution, different sources of thermal and quantum noise). 
The
original meaning of decoherence was specifically designated to
describe the loss of coherence in the off-diagonal elements of the
density operator in the energy eigenbasis\cite{breuer}. More
recently, with the additional goal of better understanding the effects
of environmental interactions on QIP and other forms of quantum
control, decoherence has been defined as non-unitary dynamics
induced by system-environment couplings\cite{zurek03, revlidardfs03}.

This non-unitary dynamics is not limited to the known relaxation
processes of dissipation and dephasing, traditionally associated
with $T_{1}$ and $T_{2}$ processes. Non-unitary or `irreversible' dynamics also includes processes in which the system
is conditioned to follow a specific quantum path by a measurement
at certain times, i.e. conditional quantum evolution or processes
in which correlations in the bath properties may play an important
role in the system's dynamics (non-Markovian effects).

Irreversible dynamics is generally seen as an undesirable consequence of QIP hardware. However, recent studies have shown
that decoherence channels can actually be exploited to favor quantum
coherence (e.g. in order to prepare two-qubit entangled
states\cite{plenio99, nicolosi04} or to perform high-fidelity
quantum gate operations\cite{beige03}). Here we briefly summarize
three approaches that give insight into how to avoid decoherence
sources, or how to use them in order to promote the persistence of quantum coherence. The first two approaches can be derived within the general
framework of processes for which all memory effects in the
environment can be discarded, i.e. Markovian dynamics which can be 
described by Lindblad-like equations\cite{breuer}. The third type of approach, in contrast, is aimed at understanding
and exploiting the bath's memory effects.

\subsubsection {\em Conditional quantum evolution}
Conditional quantum evolution, also known as the quantum jump approach, has mainly been studied in connection with optical quantum systems\cite{plenio98}. In this approach, the environment or bath functions are subject
to a continuous series of measurements which have the potential to interrupt the unitary evolution of the system at infinitesimal time intervals.
Conditioned on a particular observation record for the decay
channels, the system follows non-Hermitian dynamics in which the
effective time-evolution does not preserve the norm of the state.
This approach has been reviewed in Ref.\cite{plenio98}.

The possibility of using decay channels to prepare entangled
states have been discussed in the case of non-interacting qubits\cite{plenio99} as
well as for interacting qubits\cite{nicolosi04} that are identically coupled
to the quantized mode inside a leaky cavity. Subject to the
condition that no photon is detected, they both demonstrate that a successful
measurement performed on a sufficiently large timescale, generates an
uncorrelated state for the qubits and cavity,
leaving the two-qubit subsystem in its antisymmetric maximally
entangled state.
In the framework of this conditional evolution, it has also been shown that
dissipation can be exploited to implement fast two-qubits gates
with ground state qubits coupled to a common vibrational
mode\cite{beige03}.

A common criticism of the above schemes is that the gate operation
is probabilistic since there is always a non-zero probability of
obtaining an undesirable outcome, i.e. a photon is detected and hence
the gate operation has failed. Therefore in a broader
perspective, a primary motivation for research on conditional
quantum evolution is the prospect of developing general methods
for real-time feedback control of open quantum
systems\cite{mabuchi02}.

\subsubsection{\em Decoherence-free subspaces and subsystems}
The central idea of decoherence-free subspaces is the
identification of the dynamical symmetries in the
system-environment interaction, in order to construct a basis of collective
states that are robust to
dissipation and dephasing\cite{duan98, lidar98}.  In this approach, all qubits are
identically coupled to the same environment such that the
qubit-permutation symmetry gives rise to collective decoherence which
can be effectively overcome if the qubits are encoded in decoherence-free
states. These states have been identified as singlets
or one-dimensional irreducible representations of the algebra generating the
dynamical symmetry\cite{lidar98}. 

Following these ideas there have been proposals for inducing
subspaces which are effectively decoupled, via external time-dependent
Hamiltonians. This is known as dynamical decoupling\cite{viola02} 
and it appears to be a promising alternative for suppressing
decoherence in solid-state systems which are subject to strong low-frequency noise\cite{lidar04}. A further
generalization and unification of various schemes to avoid
decoherence is provided by the concept of noiseless subsystems\cite{zanardi01} 
in which the information can be encoded
in higher-dimensional representations. This generalization has
become the basis of a full theory of universal and fault tolerant
QIP on decoherence-free states\cite{kempe01}. A recent review of the theory of
decoherence-free subspaces and it extension to subsystems is given
in Ref.\cite{revlidardfs03}.

\subsubsection {\em Non-markovian approaches}
A particular motivation for understanding non-Markovian
effects in the dynamics of an open quantum system, comes from
experimental and theoretical studies in semiconductor nanostructures 
which have shown that the
relaxation-time approximation may be questionable in the ultrafast
optical regime within which these systems are being explored for QIP. The reason is that these approximations overestimate decay effects on the short timescales within which the system is
dynamically evolving\cite{borri01, rodriguez02, rodriguez03}.

These studies tie in to the more fundamental goal of advancing the theory of open quantum
systems, in order to develop analytical methods in which correlations in the
environment are
incorporated\cite{breuer}.  A specific goal 
is to construct evolution equations for the reduced density matrix,
that generalize the Markovian Lindblad master equation in order to include
bath memory effects and yet which remains both numerically and analytically tractable\cite{daffer03,breuer04,shabani04}. Such non-Markovian extensions are required in order to preserve complete positivity, thereby ensuring that the system dynamics is
compatible with the joint system-environment unitary evolution.

The importance of considering correlation effects within the bath, lies
in the fact that such interference effects might actually reduce the effective decay-rate for coherence. For example, it has recently been shown  that
decoherence control is possible through a Parrondo-like
effect\cite{chufan02} in which two correlated decoherence
sources are made to effectively cancel.

\setcounter{section}{2}\newpage
\section{All-optical QIP in semiconductor nanostructures}
%*****************************************************
%\subsection{Role of ultrafast optics in nanostructures}
%*******************************************************  
The possibility of
controlling the coherent dynamics of an open quantum system, is a fascinating topic
from both technological and fundamental physics perspectives.  In semiconductor
nanostructures, the availability of ultrafast lasers with their wide ranges of pulse
widths, wavelengths, pulse energies, and pulse repetition rates, has made possible the
coherent control of  matter dynamics in the {\em transient} regime, i.e. before
relaxation processes destroy the coherence created by an ultrafast optical
excitation\cite{shah}. For optically-induced polarization of a quantum dot, the
timescale of the transient regime  has been found to range from tens of
picoseconds\cite{bonadeo98} at low temperatures, down to several hundreds of
femtoseconds at room temperature\cite{borri99}. At low temperatures the mechanism of
decoherence is mainly determinded by radiative decay\cite{bonadeo98} while at higher
temperatures it is dominated by pure dephasing processes which are themselves driven by
interactions of the charge carriers with acoustic phonons\cite{krummheuer02}.

\begin{figure}
\centerline{\resizebox{8cm}{!}{\includegraphics*{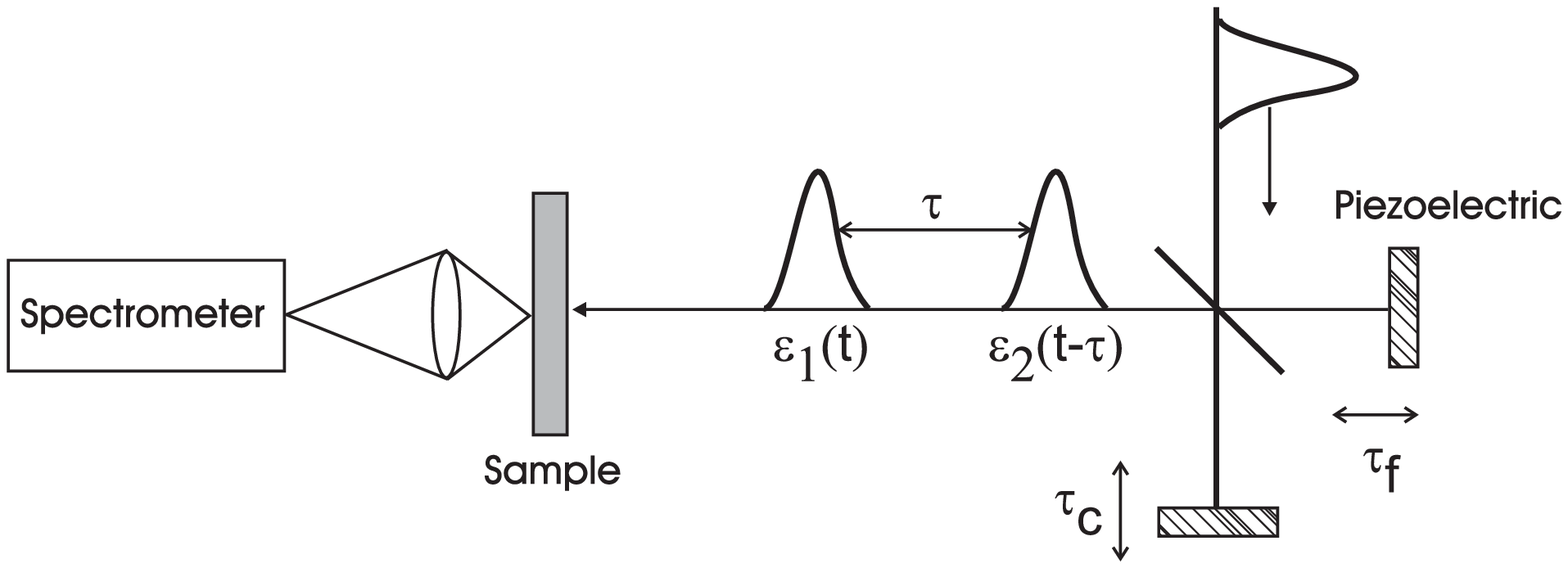}}}
\caption{Schematic diagram of the experimental setup used
for ultrafast coherent control of an exciton in a dot in
Ref.\cite{bonadeo98}. Wave interferometry is achieved by using two
phase-locked optical pulses $\varepsilon_1$ and $\varepsilon_2$
delayed a time $\tau=t_f+t_c$. The pulses interact with the QD at
times $t=0$ and $t=\tau$ creating a coherent superposition of the
dipole excited state.} \label{fig:setup}
\end{figure}

Unprecedented levels of coherent control have been demonstrated on superpositions of
exciton and biexciton states in a single quantum dot, using ultrafast
optics\cite{chen00,stievater01,chen02,steel03} (see figure \ref{fig:setup}). Earlier experiments showed coherent
manipulation of the exciton wave function\cite{bonadeo98}, coherent evolution between
two different excitonic states with orthogonal polarizations\cite{chen00}, as well as
Rabi oscillations between the vacuum state of excitons and a linearly polarized
excitonic state\cite{stievater01}. Furthermore quantum superpositions have been
reported between the ground state and the biexciton state\cite{chen02} as well as
exciton-to-biexciton Rabi oscillations\cite{steel03}. These experimental achievements
have been integrated in a set-up to perform an all-optical quantum gate in a single
quantum dot\cite{steel03}. Although this system is not scalable by simply adding
arbitrary numbers of excitons into a dot, it demonstrates the potential for ultrafast
optically-driven manipulation in scalable architectures based on multidot arrays.

Proposals for large-scale QIP in semiconductor QDs have been made using two different
realizations of the qubit: the excess electron spin\cite{loss98, bukard99} and an
electron-hole pair excitation or exciton\cite{quiroga99, reina00, biolatti00,
rinaldis02, lovett03}. The former qubit benefits from  a weak coupling to the
environment, yielding longer coherence times that are of the order of
microseconds\cite{kikkawa98}. However it lacks a fast and easy tunability of coupling
between qubits. The latter candidate provides an extremely fast interaction between
qubits, of the order of picoseconds for several proposals, but it suffers from short
dephasing times.  In order to achieve a high ratio between coherence and gate operation
times, hybrid schemes which merge these various advantages have recently been
proposed\cite{piermarocchi02, pazy03,calarco03, nazir04}. In particular, these latter
approaches use optical excitations to control  the coupling between QD spins. Schemes
based on spin-flip Raman transitions mediated by a high-Q semiconductor microcavity,
have also been proposed\cite{imamoglu99, imamoglufp00, feng03}.

The common goal of these approaches is to use ultrafast technology to
achieve all-optical control of the qubit state and inter-qubit interactions in
semiconductor QDs, and to then scale up this procedure to produce large-scale QIP. In
this section we review proposals aimed at this end.

\subsection{Ultrafast schemes with excitons as qubits} 
\begin{figure}
\centerline{\resizebox{8cm}{!}{\includegraphics*{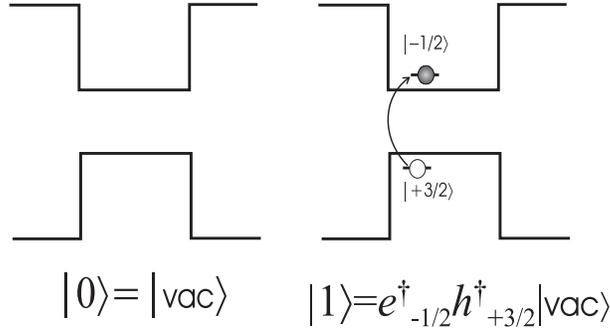}}}
\caption{Logical states of an excitonic qubit.  $|0\rangle$ denotes vacuum
  of excitations and $|1\rangle$ referes to the presence of a ground-state exciton made 
up of a hole with angular momentum $+3/2$ and an electron with angular momentum
$1/2$.}
\label{fig:exciton}
\end{figure}

Optically-driven semiconductor
QDs lie at the heart of many QIP proposals\cite{quiroga99, reina00, biolatti00, rinaldis02, lovett03}. These systems offers discrete energy levels
which can be quickly and accurately addressed in space, and can be built
using well-established semiconductor fabrication technology. The natural qubit in the
QD system is given by the absence ($|0\rangle$) and presence ($|1\rangle$) of a
ground-state  electron-hole pair, the so-called exciton as illustrated in figure
\ref{fig:exciton}. Experimental evidence suggests that the decoherence time of this
qubit is mainly limited by the radiative lifetime\cite{bonadeo98}.

\begin{figure}
\centerline{\resizebox{8cm}{!}{\includegraphics*{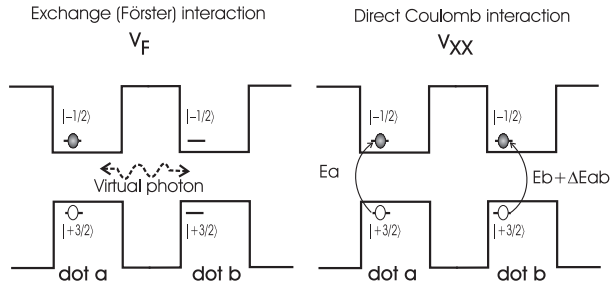}}}
\caption{Schematic illustration of interdot interactions. Resonant energy 
transfer(F\"orster) 
process thereby an exciton is destroyed in one quantum dot and
re-created on the other quantum dot via exchange of a virtual photon(left side). 
When each dot contains an exciton, direct electrostatic
Coulomb interaction between quantum dots takes place. This lead
to individual energy shifts which in presence of a static electric field
in the $xy$ plane become significant, and can be exploited to generate
an energy selective two-qubit gate (right side).}
\label{fig:interdot}
\end{figure}

Several QIP schemes have proposed exploiting exchange and/or direct Coulomb interactions
between spatially separated excitonic qubits in coupled QD
structures\cite{quiroga99,reina00, biolatti00,lovett03}. The Coulomb
exchange interaction in QD molecules gives rise to a non-radiative resonant energy
transfer (i.e. {\em F\"orster} process) which corresponds to the exchange of a virtual
photon, thereby destroying an exciton in a dot and then re-creating it in a close by dot (see left-hand side of figure
\ref{fig:interdot}).  Quiroga and Johnson\cite{quiroga99} and Reina et
al.\cite{reina00} have discussed how to exploit the F\"orster interaction to prepare
both Bell and Greenberger-Horne-Zeilinger (GHZ) entangled states of excitons, using far-field
light excitation to globally address two and three quantum dots in a spatially
symmetric arrangement. The Hamiltonian describing the formation of single excitons
within the individual QDs and its interdot F\"orster hopping is given by\cite{quiroga99}
\begin{equation}
\mathcal{H}(t)=H_0+ H_F +  H_{ext}(t)\ .
\end{equation} 
Here the single-exciton Hamiltonian is given by
\begin{eqnarray} H_0=\frac{\varepsilon}{2}\sum_{l=1}^N(e^{\dag}_l e_l - h_l
h_l^{\dag})\ ,
\end{eqnarray} 
the interdot F\"orster interaction can be written as
\begin{eqnarray} H_F=-\frac{V_F}{2}\sum_{l,l'}^N (e^{\dag}_l h_{l'} e_{l'} h_l^{\dag} +
 h_l e^{\dag}_{l'}h^{\dag}_{l'}e_l)\ ,
\end{eqnarray}
the coupling of the carrier system with a classical laser of
amplitude
$E(t)$ is described by
\begin{eqnarray} H_{ext}(t)=E(t)\sum_{l=1}^N e^{\dag}_l h^{\dag}_l + E^*(t)\sum_{l=1}^N
h_l e_l
\end{eqnarray} 
and all constant energy terms have been ignored. The electron (hole)
creation operator in the $l$th QD is designated by
$e^\dag(h^\dag)$, $\varepsilon$ is the QD band gap and $V_F$ denotes the strength of
the F\"orster coupling between dots. The equidistant configuration of the multidot
system favors the definition of global quasispin operators, which in turn enable an
understanding of the dynamics of the multiexciton system in terms of the dynamics of its
associated global quasispin\cite{quiroga99}.
 Starting with the initial condition of a vacuum of excitons, the preparation of a Bell state such as $|\Psi\rangle=\alpha|00\rangle + \beta|11\rangle$ where $|11\rangle$ denotes the simultaneous presence of two excitons in a double dot structure, follows from the application of a finite rectangular pulse of the form
$E(t)=A cos(\omega t)$ with central frequency $\omega$ and sub-picosecond duration. When
the multidot system is arranged in a linear configuration where exciton hopping takes
place only between nearest neighbors, the Hamiltonian $\mathcal {H}(t)$ takes the form
of the one-dimensional $XY$ Heisenberg model\cite{zhang03}. The derivation of the
effective Hamiltonian relies on introducing the following local $1/2-$pseudospin
operators for electron-hole pairs in each dot:
\begin{eqnarray}
\sigma_l^x &=&\frac{1}{2}(e^{\dag}_l h^{\dag}_l + h_l e_l )\nonumber\\
\sigma_l^y &=&\frac{-i}{2}(e^{\dag}_l h^{\dag}_l - h_l e_l )\\
\sigma_l^z &=&\frac{1}{2}(e^{\dag}_l e_l - h_l h_l^{\dag})\nonumber\ \ .
\end{eqnarray} 
Hence without the optical field term, the effective $XY$ Hamiltonian
reads
\begin{eqnarray}
\mathcal{H}_{eff}(t)=\varepsilon \sum_{l=1}^N \Delta \sigma^z_l-V_F\sum_{l=1}^{N-1}
(\sigma_l^-\sigma_{l+1}^+ + \sigma_l^+\sigma_{l+1}^-)
\end{eqnarray} 
where $\Delta$ is the detuning from the resonant excitation and
$\sigma_l^{\pm}=\sigma_l^x \pm i\sigma_l^y$. An interesting point which emerges from
this analysis is the difference of timescales required to generate entanglement in
the different configurations. In particular, it turns out that GHZ states require
longer times in a symmetric multidot arrangement than in a linear
configuration\cite{zhang03}. It would be interesting to check whether such a statement
also applies to a more general QIP protocol.

The direct electrostatic interaction between two excitons (see right-hand side of
figure 
\ref{fig:interdot}) changes the energy of the excitonic transition, and is known
as a `bi-excitonic' shift\cite{jacak}. In  the presence of an in-plane electric field
the excitons acquire a permanent electric dipole and the bi-excitonic shift becomes
significant\cite{biolatti02}. This electrostatic dipole-dipole interaction has been
proposed as a physical mechanism for implementing a controlled-NOT operation in a
double dot structure, as discussed by Biolatti et al.\cite{biolatti00}. Assuming that
the distance between the dots is large  enough to prevent single-particle tunnelling
but at the same time that they are sufficiently close to assure a strong interdot
Coulomb interaction, the effective Hamiltonian governing the dynamical evolution of the
system can be written as\cite{biolatti00}
\begin{eqnarray}
\mathcal{H}_{eff}(t) =  H_0 + H_{xx}=\sum_{l=a,b}E_l \hat n_l +
\frac{1}{2}\sum_{l\neq l'}\Delta E_{ll'}\hat n_l \hat n_{l'}
\end{eqnarray} where $\hat n_l$ denotes the excitonic occupation number operator with
eigenvalues $0$ and $1$, with $0$ and $1$ denoting the absence and presence
of an exciton in the $l-$th QD respectively. Here $E_l$ denotes the ground-state
exciton energy;
$\Delta E_{ll'}$ is the bi-excitonic shift in the presence of an electric field, which
only arises if the qubits in  dots $l$ and $l'$ are in state
$|1\rangle$ as illustrated in right-hand side of figure \ref{fig:interdot}.  This state-dependent interaction can be exploited to design conditional
operations with properly adjusted two-color laser pulses. For example in the case of
two coupled dots $a$ and $b$, the transition $|1_a0_b\rangle\mapsto
|1_a1_b\rangle$ could be achieved on the sub-picosecond timescale by the following
sequence : First apply a $\pi$ rotation of the state of qubit $a$ ($|0\rangle \mapsto
|1\rangle$) and then apply a second pulse with frequency $E_b+\Delta E_{ab}$ to perform
a $\pi$ rotation of qubit $b$. We note that this analysis neglects the presence of the 
F\"orster interaction. 
In addition, one of the possible difficulties with this scheme is the need for
an external electric field, since this would require the presence of  electrical
contacts which not only increase the complexity of the set-up but would also imply an
additional source of decoherence due to electromagnetic fluctuations. In order to
circumvent this requirement, the same group\cite{rinaldis02} have proposed the use of
QD structures with  built-in electric fields, as observed in GaN dots.  

The interplay between the resonant energy transfer ($V_F$) and the interdot bi-exciton
binding energy ($\Delta E_{ll'}=V_{xx}$) has been studied numerically by
Lovett et al.\cite{lovett03, lovett03-2}. These authors have shown that by taking into
account both the F\"orster interaction and the bi-exciton binding energy, it is possible
to develop an  energy-selective approach to prepare entangled states of excitons
and to perform the CNOT operation in QD molecules. They consider two QDs
$a$ and $b$ having different sizes, which implies that in  the absence of interdot interactions
there is an energy difference between the excitonic transitions in the dots denoted by
$\Delta_0$.  Hence two regimes, determined by the ratio $V_F/\Delta_0$, can be
explored.  When the F\"orster coupling is dominant ($V_F/\Delta_0
\gg 1$) and the initial state is $|0_a1_b\rangle$ which denotes a single exciton in dot $a$
and no-exciton in dot $b$, the system can naturally evolve into a singlet which
is maximally entangled state. When
$V_F/\Delta_0 \ll 1$, the bi-exciton binding energy $V_{xx}$ becomes dominant and a
{\footnotesize CNOT} operation can be
implemented which is driven by pulses of different frequencies. In this case, entangled
states can also be prepared by a properly designed laser-pulse sequence starting from the vacuum of excitons.

Inter-dot interactions in the presence of interband excitations, have recently been
described in the context of a multipolar Quantum Electrodynamics (QED)
Hamiltonian\cite{sangu04}.  This treatment allows one to understand the previously
discussed interactions  in terms of the exchange of transverse photons and electrostatic
contributions, but also points out the physical mechanism to induce dipole-forbidden
transitions that are mediated by an optical near-field\cite{sangu04}.

All these schemes share the advantage of providing fast two-qubit operations within
the sub-picosecond time scale, as a result of resonant energy transfer or the
interdot-biexciton binding energy. However this timescale is still comparable with
the dephasing time of the  exciton dipole. A good question at this point is then: {\em
is it possible  to take advantage of this ultrasfast interaction, and hence
ultrafast technology itself, to develop a combined scheme which integrates a long
coherence qubit with fast two-qubit operations?} This is precisely the question that
recent proposals have been exploring\cite{piermarocchi02, pazy03,calarco03, nazir04}. In
what follows, we summarize proposals that share the idea of exploiting electron-hole
excitations to control the coupling between QD spin qubits.

\subsection{Exciton-assisted spin-based quantum computation} 
Single electron spins
confined in quantum dots have been proposed as qubits \cite{loss98, bukard99} because
of their long relaxation times which are well into the microsecond
timescale\cite{kikkawa98}. It has been suggested that the interaction between spins
could be controlled by electronic gates\cite{loss98}. However an all-optical approach
is clearly more desirable since this would avoid the additional fabrication of gates and
the unavoidable fluctuations in electromagnetic fields arising from these gates.
As for the other all-optical approaches which have formed the focus of
this article, it instead makes sense to try to exploit the major advances in ultrafast
laser technology which continue to be developed\cite{shah}. Indeed, it has already been
demonstrated experimentally that a single electron spin in a QD can be 
probed optically\cite{gupta01}. 

\begin{figure}
\centerline{\resizebox{8cm}{!}{\includegraphics*{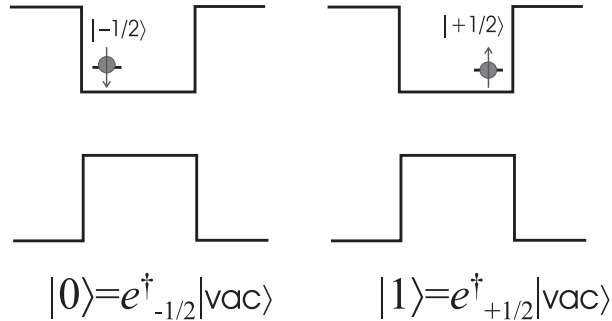}}}
\caption{Logical states associated to the spin of a single
electron confined in a QD.} \label{fig:qdspin}
\end{figure}

The idea behind spin-based QIP assisted by excitons\cite{piermarocchi02,
pazy03,calarco03, nazir04} is to exploit virtual or resonant interband excitations to
optically induce and control interactions among spin-qubits which are localized in
different QDs. In these schemes the logical qubit is defined by the spin-states of a
single conduction-band electron confined in a QD:
$|0\rangle=|m_z=-1/2\rangle$ and $|1\rangle=|m_z=1/2\rangle$ as shown in figure
\ref{fig:qdspin}.

\begin{figure}
\centerline{\resizebox{8cm}{!}{\includegraphics*{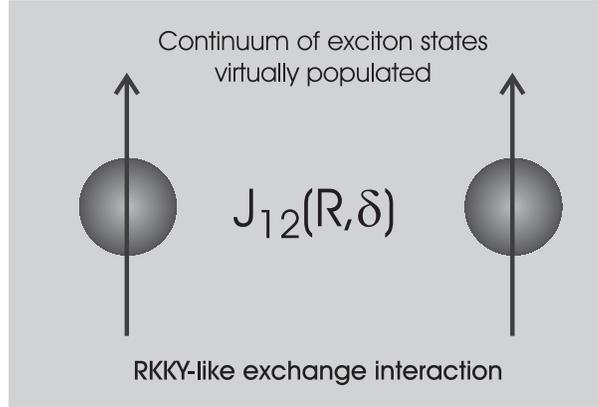}}}
\caption{Schematic illustration of the effective optical RKKY
exchange interaction between electrons in separated quantum dots
via an intermediate continuum of exciton states produced by an
off-resonance optical excitation. Here $\delta$ is the detuning
between the laser frequency and the band gap frequency in the host
material, $R$ is the distance between confined electrons.}
\label{fig:rkky}
\end{figure}

The first scheme which used exciton states to induce indirect exchange interactions
between electron spins localized on different QDs, was discussed by  Piermarocchi et
al.\cite{piermarocchi02}. The excess electrons confined in spatially separated
QDs, interact with a delocalized electron-hole pair which has been excited by an
optical pulse which is itself detuned  with respect to the continuum of exciton states
in the host material.  Keeping only the lowest-order contribution, exchange Coulomb
interactions between the localized and the optically-excited conduction electrons lead
to an effective spin-spin exchange coupling between QD spins of the type discussed in
the previous section, i.e. 
$H_{eff}=-2J_{12}S_1S_2$ with $S_{i=a,b}$ being the electron spin in dot $i$. In this
scheme the coupling parameter $J_{12}$ is always positive (ferromagnetic interaction)
and depends on the detuning frequency between the laser and the excitonic
transition in the continuum, and the interdot separation as illustrated in figure \ref{fig:rkky}. This coupling mechanism is
analogous to a Ruderman-Kittel-Kasuya-Yosida (RKKY) interaction between two well
separated magnetic impurities mediated by either conduction electrons or
excitons\cite{demelo95}, except that in the present case the intermediate electron-hole
pair is produced by an external optical field.

\begin{figure}
\centerline{\resizebox{8cm}{!}{\includegraphics*{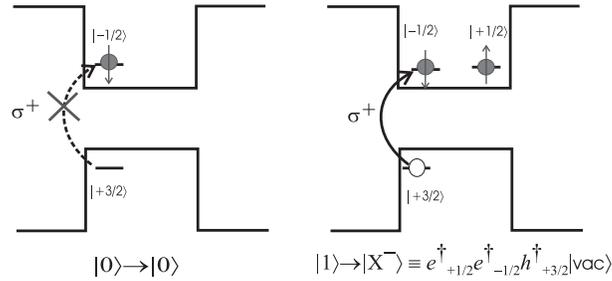}}}
\caption{Optically controlled Pauli-blocking mechanism. If the
spin qubit is in $|0\rangle$ a creation of an exciton with
electron angular momentum $-1/2$ is inhibited (left side). If the
qubit is in $|1\rangle$, then nothing forbids the creation of such
exciton. If the qubit is in a coherent superposition of its
logical states, then it is transformed into a charge
superposition: $|0\rangle +|1\rangle\mapsto
|0\rangle+|X^-\rangle$.} \label{fig:pauli}\end{figure}

\begin{figure}
\centerline{\resizebox{8cm}{!}{\includegraphics*{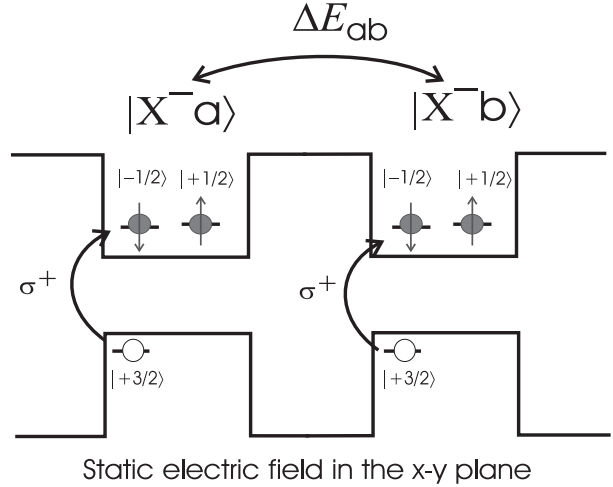}}}
\caption{Bi-excitonic shift. Interaction between QD spins is
contolled by the electrostatic Coulomb interaction between
optically created trion states in dots $|x^-_{a}\rangle$ and
$|x^-_{b}\rangle$. In presence of a static electric field the
trion states acquire permanent dipoles while the Coulomb
interaction induces a significant energy shift $\Delta E_{ab}$ in
exciton states. In combination with the Pauli-exclusion mechanism,
this energy shift can be exploited to perform a quantum phase
gate.} \label{fig:biexc}
\end{figure}

The second scheme we discuss is based on conditional spin and exciton dynamics and
employs a Pauli-blocking mechanism\cite{pazy03, calarco03}. As illustrated in figure
\ref{fig:pauli}, when the QD is excited with left-handed circularly polarized light the
presence of an excess electron in its spin-down state ($|0\rangle$) inhibits the
creation of an exciton with electron angular momentum $m_z^e=-1/2$. By contrast
if the qubit is in its spin-up state ($|1\rangle$) then nothing prevents the creation of
the electron-hole pair. This yields a state-selective coupling of
the logical states $|0\rangle$ and
$|1\rangle$ to the auxiliary state, defined as $|x^-\rangle$, which describes a QD with
two electrons with opposite spins occuping the same energy level together with a single
hole.
This scheme also exploits the Pauli exclusion principle and electrostatic
exciton-exciton interactions in order to implement a two-qubit phase gate. A
rotation, given by an accumulated phase $\theta$, only occurs when both qubits are
in their logical state $|1\rangle$:
\begin{eqnarray} |m,n\rangle \mapsto e^{i\theta m n}|m,n\rangle \quad \quad m \;
 \epsilon \{0,1\}\ .
\end{eqnarray} 
As discussed by Calarco et al\cite{calarco03}, the key mechanism here is
the energy shift due to the electrostatic dipole-dipole interaction between excitons
when an external static electric field is applied (see figure \ref{fig:biexc}). As
mentioned earlier, this bi-excitonic shift occurs  only in the case when both qubits are
in their spin-up state. The effective Hamiltonian describing the situation for two
adjacent QDs ($a$ and
$b$) is given by\cite{calarco03}
\begin{eqnarray}
\mathcal{H}_{eff}=\sum_{\nu=a,b}\left
(\frac{\Omega_{\nu}(t)}{2}|x^-\rangle_{\nu}\langle 1| + H.c.\right )- \Delta
E_{ab}|x^-\rangle_{a}\langle x^-|\otimes |x^-\rangle_{b}\langle x^-|\ \ .
\end{eqnarray} 
Here $\Omega(t)$ is the effective Rabi frequency between the
single-electron state $|1\rangle $ and the trion state
$|x^-\rangle$. In this proposal, as in most of the spin-based schemes, realizing
the optical rotation of a single spin represents a significant challenge. To
overcome this difficulty, the authors suggest the implementation of Raman transitions
via light-hole levels
$m_z^h=\pm 1/2$ in situations where these states are the hole ground-states,
a situation which does occur in II-VI semiconductors nanostructures.

A modification of the above scheme to achieve the spin-couplings via inter-dot
resonant or F\"orster energy interaction ($V_F$) has been discussed in
Ref.\cite{nazir04}. This modified scheme requires the excitation of a single
ground-state exciton instead of interdot-biexciton states.

\subsection{QIP schemes with microcavities and quantum dots} 
As discussed in the
previous section, previous work on conditional quantum dynamics has demonstrated that
cavity QED could play a key role in future quantum communication or computation
schemes. More specifically, it has been shown that cavity-mediated interactions provide
a means for preparing entangled states, in addition to providing a means
for transmitting quantum information
 between distant nodes in a quantum network. Following this line of thought, Imamoglu
et al.\cite{imamoglu99, imamoglufp00} proposed a system which relies on the use of a
quantized cavity mode and applied laser fields in order to mediate the interaction
between spins of distant, doped QDs. The central idea in this scheme is to implement
cavity assisted spin-flip Raman transitions and to couple pairs of qubits via virtual
photons in the common vacuum cavity mode.

Assuming that a uniform magnetic field is applied along the $x$ direction,
the QD qubit is defined by the spin states
$|m_x=-1/2\rangle=|0\rangle$ and $|m_x=1/2\rangle=|1\rangle$ of the single
conduction-band electron. The scheme's proposers consider QD spins interacting with
a
$x-$polarized vacuum cavity mode and a
$y-$polarized laser field in order to implement cavity-assisted spin-flip Raman
transitions between the two spin states, in close analogy with atomic cavity-QED
schemes\cite{pellizzari95}. By adjusting the frequencies of the individual lasers in
order to establish a near two-photon resonance condition, such that the cavity is
virtually excited, the following effective Hamiltonian can be obtained:
\begin{eqnarray} H_{int}=\sum_{i\neq j}\frac{\tilde
g_{ij}(t)}{2}[\sigma_{10}^i\sigma_{01}^j +\sigma_{10}^j\sigma_{01}^i]\ \ .
\end{eqnarray} 
Here $\sigma_{10}^i=|1\rangle\langle0|$ is the spin projection operator
for the $i$th QD, and $\tilde g_{ij}(t)$ corresponds to the product of the two-photon
coupling coefficients for the spins in QDs $i$ and $j$. It has been shown that the
two-qubit coupling between any pair of QDs  can be carried out in parallel and on
sub-nanosecond timescales.

A modification of the above scheme so as to combine Pauli-blockade effects with the
microcavity scheme has been discussed in Ref.\cite{feng03}. In this  modified scheme,
the quantum information is defined by the  states $|m_z=\pm 1/2\rangle$ of the single
conduction electron. The cavity mode, which is assumed to be right-hand polarized,
induces an electron-hole pair excitation in a dot only if the excess spin electron is
in the state
$|m_z=-1/2\rangle$. Individual lasers are assumed to be linearly polarized, and
have their frequencies adjusted such that the Raman transition between
the spin states can occur.

\subsection{Decoherence control through optical pumping} Despite its atom-like
properties, there are fundamental features that distinguish a quantum dot from an atom.
One of these is the completely different role that hyperfine interactions play in the
two quantum systems. In contrast to valence electrons of an atom, a single electron in a
dot is confined on a lengthscale which extends over many lattice sites. Consequently,
the electron spin interacts with a few thousand randomly-oriented nuclear spins.
Interactions with this unpolarized bath present a  decoherence mechanism for an
electron spin in a quantum dot\cite{khaetskii02}. Imamoglu et al.\cite{imamoglu03} have
discussed an optical scheme to suppress this decoherence mechanism. The main idea is to
use hyperfine interactions to polarize the nuclear field. This is achieved by shifting
the energy of the initial spin-up electronic state via the ac-Stark effect in order to
create a resonant condition for the electron-nuclear spin-flip transition.

Interestingly, Shabaev et al. \cite{shabaev03} have shown that even in the presence of
random hyperfine interactions with nuclear spins, a strong resonant optical excitation
of the electron spin to an intermediate trion state provides readout capability of the
spin state during the relaxation process of the trion state\cite{shabaev03}. Combined
with a permanent transverse magnetic field, such an optical excitation also provides  a
way to initialize the spin into a state with a well-defined phase\cite{shabaev03}.

\subsection{Concluding remarks} Optical properties of semiconductor QDs can be tailored
by varying the size, shape and composition material of the QD, thereby offering a
suitable scenario to implement all-optical approaches for the coherent control of
qubits and their interactions.  Semiconductor nanostructures integrated with ultrafast
optics technology,  are therefore an attractive solid-sate alternative for constructing
scalable and fault-tolerant architectures in order to implement quantum computation and
communication, as well as quantum simulation protocols.

There is still a very long way to go before  large-scale quantum processors can be made
out of  QD arrays. Indeed the state of the field is such that it would be a major
scientific breakthrough if someone were to demonstrate quantum entanglement
involving just a few QDs, let alone control or manipulate this entanglement. However
there are many experimental groups trying to do exactly that -- and eventually
someone will manage. Along the way, there are many open
problems which will need to be solved,  and several new research themes will emerge.
These open problems include a deeper understanding of decoherence mechanisms and
readout for a single electron-spin, and experimental signatures of  superposition and
entanglement for excitonic and spin qubits and for the entanglement between a QD
system and photons. In the next section we discuss some of the trends we forsee for
future developments.

\setcounter{section}{3}\newpage
\section{Trends for future developments}

The study of QIP in nanostructures has opened up many new
questions. As a result of the presence of similar physical mechanisms of
interactions in organic and semiconductor systems, one such question concerns the
extent to which organic and biomolecular systems offer a viable alternative
for QIP, e.g. `bio-QDs' such as the LHI and LHII complexes used by Nature to
harness the energy of a photon in photosynthesis. A second set of questions
relates to how one can best exploit the interaction between non-classically
correlated photons and nanostructures. We address each of these briefly.

\subsection{QIP in organic and biomolecular nanostructures} 
Quantum dots
embedded in organic structures offer novel physical properties. In particular
they allow the formation of exciton states which exhibit large oscillator
strengths and strong coupling to the light. These properties yield a large
coherence length and high optical nonlinealities\cite{huong00} which may be
exploited for QIP.

It has also been recognized that the coupling mechanisms occurring in
QD molecules have the same physical origin as those present in existing natural
systems, such as the light-harvesting complexes (LHI and LHII) for which
excitonic interactions and energy transfer processes play a central
role\cite{revhu02}. At nanosecond timescales, the fluorescence resonance energy
transfer (FRET) which is associated with the F\"orster process in
light-harvesting complexes and in FRET-coupled dye pairs, exhibits an incoherent
dynamics. However, coherent FRET signals might possibly be found at
pico- and femtosecond timescales, as discussed earlier for coupled QDs.
Due to these basic similarities between inorganic QDs and biomolecular
nanostructures, novel ideas for using these latter natural structures to process
quantum information are beginning to be explored\cite{lovett03,sekatskii03}.

From our own perspective, we believe that {\em hybrid} bio-nano QIP 
systems will emerge as an
important field of study in the future. This inter-disciplinary field will need
to combine novel ideas and understanding from the physical, chemical and
biological sciences, as well as the biotechnology and nanofabrication
industries. Whether fully quantum, mixed quantum-classical, or just classical
devices can be built, remains to be seen. However all three prospects are
exciting, whether it be classical information processing (IP) or full quantum
information processing (QIP) systems which finally emerge.

\subsection{Nanostructures and entangled photons}

We have mentioned that one can take advantage of the F\"orster interaction
between two quantum dots, which are globally addressed by a laser beam, in order
to generate Bell states of the type of
$\alpha |00\rangle + \beta |11\rangle$, where the state $|11\rangle$ indicates
the simultaneous presence of two excitons in the double-dot
array\cite{quiroga99}. This indicates that the formation of such entangled
states can be achieved through a two-photon excitation. In fact, it has been
experimentally shown that the coherent resonant dipole interaction between
molecules can be probed via a two-photon transition in which both molecules are
simultaneously excited
\cite{hettich02}. An interesting question therefore arises as to what
kind of quantum interference phenomena occur when the molecular (or QD) pair
is excited by two photons which are entangled. Sources of entangled pairs of
photons in the visible spectrum, are now available\cite{volz01}. In fact sources
of three and four entangled photons have very recently been
reported\cite{walther04, mitchell04}, which in turn allows one to extend the
question beyond two molecules (or QDs) to many-particle systems. This is also an
exciting area for future study, both theoretically and experimentally.

\subsection{Photon statistics and dynamics of non-classical correlations}

In quantum systems with optical outputs, it is expected that experimental
signatures of entanglement take the form of non-classical statistical
correlations of the emitted light\cite{olaya01}. In matter-light coupled
systems, photon statistics has already proved to be a valuable tool for the
identification of quantum signatures such as photon-antibunching in the
resonance fluorescence of a two-level atom or a single quantum
dot\cite{michler00}. Indeed, second-order photon correlations are expected to
exhibit signatures of coherent superpositions in single and double QD
nanostructures\cite{rodriguez02,rodriguez03}. In a parallel
development, Hanbury-Brown-Twiss experiments have been performed to measure
intensity correlations in the nonlinear response of strongly and coherently
coupled molecules\cite{hettich02}, as well as in the FRET in coupled dye 
pairs\cite{berglund02}. These
experimental achievements suggest that photon statistics could be used to
characterize the interactions in coupled quantum dots. However further
theoretical studies are required to investigate the precise details of such a
characterization.

\section{Acknowledgements}
We are extremely grateful to Luis Quiroga for useful discussions and critical reading of this manuscript. We acknowledge
funding from  the Clarendon Fund and ORS (AOC), and the DTI-LINK project
`Computing at the Quantum Edge' (NFJ).

\newpage
%\newpage
%\section{figures}
\setcounter{section}{3}

%%%%%%%%%%%%  Fig 3.1%%%%%%%%%%%%%%%%%%%%%%%%%%%%%%%%%%
%%%%%%%%%%%%%%%%%%%%%%% End fig 3.1%%%%%%%%%%%%%%%%%%%%%%%%%%%%%%%%%

%%%%%%%%%%%%%%%%%%%%%%%  Fig 3.2%%%%%%%%%%%%%%%%%%%%%%%%%%%%%%%%%%
\begin{figure}
\centerline{\resizebox{8cm}{!}{\includegraphics*{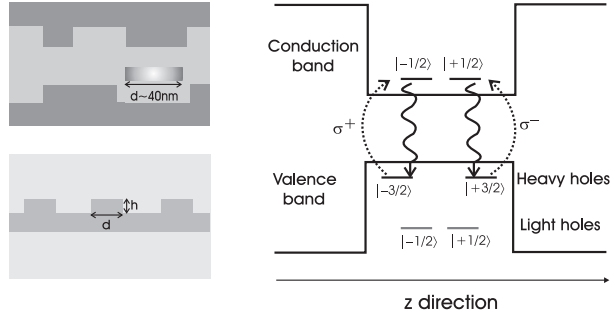}}}
\caption{Left pannel:Shematic diagrams of quantum dots. An island formed due to fluctuations in a
quantum well width (left top) and self-assembled quantum dots
grown by the Stranski-Krastanow process(left botoom). The base diameter $d$ and the height $h$
are in the range of $20-40nm$ and $3-6nm$ respectively, dedepending on the growing conditions. These latter dots
offer high potential for scalable architectures for QIP.
Right pannel: Relevant energy levels of a $III-V$ semiconductor quantum dot. Strong
confinement is assumed to be in the $z$ direction. Here $\sigma^{\pm}$ indicates right(left)-hand
circularly polirized light.}
\label{fig:qdot}
\end{figure}
%%%%%%%%%%%%%%%%%%%%%%% End fig 2 %%%%%%%%%%%%%%%%%%%%%%%%%%%%%%%%%

%%%%%%%%%%%%%%%%%%%%%%%  Fig 3.3%%%%%%%%%%%%%%%%%%%%%%%%%%%%%%%%%%
%%%%%%%%%%%%%%%%%%%%%%% End fig 3 %%%%%%%%%%%%%%%%%%%%%%%%%%%%%%%%%

%%%%%%%%%%%%%%%%%%%%%%%  Fig 3.4%%%%%%%%%%%%%%%%%%%%%%%%%%%%%%%%%%
%%%%%%%%%%%%%%%%%%%%%%% End fig 1 %%%%%%%%%%%%%%%%%%%%%%%%%%%%%%%%%

%%%%%%%%%%%%%%%%%%%%%%%  Fig 3.5%%%%%%%%%%%%%%%%%%%%%%%%%%%%%%%%%%
%%%%%%%%%%%%%%%%%%%%%%% End fig 1 %%%%%%%%%%%%%%%%%%%%%%%%%%%%%%%%%

%%%%%%%%%%%%%%%%%%%%%%%  Fig 3.6%%%%%%%%%%%%%%%%%%%%%%%%%%%%%%%%%%
%%%%%%%%%%%%%%%%%%%%%%% End fig 1 %%%%%%%%%%%%%%%%%%%%%%%%%%%%%%%%%

%%%%%%%%%%%%%%%%%%%%%%%  Fig 3.7%%%%%%%%%%%%%%%%%%%%%%%%%%%%%%%%%%
%%%%%%%%%%%%%%%%%%%%%%% End fig 1 %%%%%%%%%%%%%%%%%%%%%%%%%%%%%%%%%

%%%%%%%%%%%%%%%%%%%%%%%  Fig 3.8%%%%%%%%%%%%%%%%%%%%%%%%%%%%%%%%%%
%%%%%%%%%%%%%%%%%%%%%%% End fig 1 %%%%%%%%%%%%%%%%%%%%%%%%%%%%%%%%%

%%%%%%%%%%%%%%%%%%%%%%%  Fig 3.9%%%%%%%%%%%%%%%%%%%%%%%%%%%%%%%%%%
\begin{figure}
\centerline{\resizebox{8cm}{!}{\includegraphics*{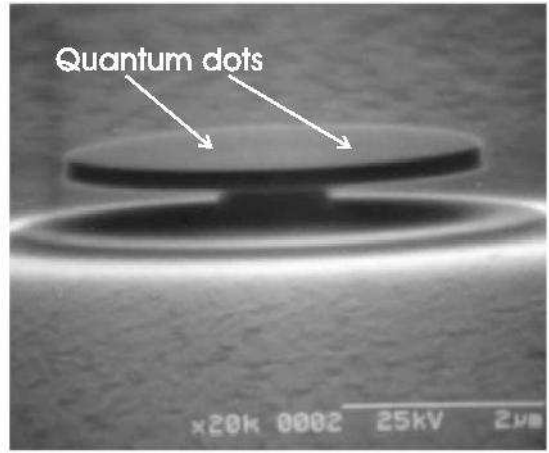}}}
\caption{Scanning electron micrograph of the GaAs microdisk nanostructure reproduced from Ref.\cite{kiraz03} with permission of the Institute of Physics Publishing \copyright 2003. The diameter of the cavity is $4.5\mu m$ with
a corresponding cavity-mode volume of $V_{cav}\simeq
200(\lambda/2n)^3$. The highest value of Q measured in this
structure exceeded 1800. The cavity contains InAs QDs at 
locations fixed during the growth. The density of QDs is  
$2\times 10^{6}$cm$^{-2}$.} \label{fig:cavity}
\end{figure}
%%%%%%%%%%%%%%%%%%%%%%% End fig 9 %%%%%%%%%%%%%%%%%%%%%%%%%%%%%%%%%

\end{document}